\documentstyle[11pt,newpasp,twoside,epsf]{article}
\def\edcomment#1{\iffalse\marginpar{\raggedright\sl#1\/}\else\relax\fi}
\reversemarginpar

\begin{document}
\title{Pulsar Kicks: Spin and Kinematic Constraints}
 \author{Roger W. Romani}
\affil{Dept. of Physics, Stanford University,
Stanford, CA 94305-4060}
%\author{Ima Co-Author}
%\affil{The Name of My Institution, The Full Address of My Institution}

\begin{abstract}
It has been noted that the Crab and Vela pulsar proper motions 
lie along the symmetry axes of their wind nebulae. In an effort to promote
this observation to a serious test of kick physics, we are using CXO images
and other data to estimate the angle between the proper motion and 
PWN (i.e. spin) axis for a number of pulsars. Here we give a progress
report on this work and the constraints that these data provide on
kick models. Present data suggest that a kick duration of $\tau_K \sim 3$s
is sufficient to explain the alignment of most pulsars. This rules out
E-M and hydrodynamic kick models, but is fairly consistent with proposed
anisotropic $\nu$ emission. However, some objects, especially PSR J0538+2817
show such good alignment that even $\nu$ models are challenged.
\end{abstract}

\section{Introduction}

	Typical pulsar velocities of $\sim 500$km/s represent a lot of momentum,
and the nature of the kick that gives neutron stars such speeds has long been
one of the major problems in compact object physics. The 
distribution of kick speeds has important implications for the observed pulsar
population, especially those in binaries; thus measurement of pulsar proper 
motion distributions and application to binary modeling sums has been a major 
activity (see Podsiadlowski; Burgay; Dewi The {\it etc.}, these proceedings).
However, ${\vec v}$ is a vector quantity and comparison of its orientation
with respect to that other relic of neutron star birth ${\vec \Omega}$,
promises to provide additional insight into the kick physics.

	The letter of Spruit \& Phinney (1998) was influential in promoting
thinking about the spin-kick connection. These authors, in fact, hypothesized
that neutron star initial angular momentum was small due to strong core-envelope coupling
in pre-collapse stars.  They suggested that an off-center kick, at impact 
parameter $d=R {\rm sin}\psi_{kick}$ imposed while the bloated proto-NS has 
radius $\sim 3 \times 10^6$cm produces a spin of
$$
\Omega_{rms} \approx 42 {\rm s^{-1}} 
\left ( {{{\rm sin}\psi_{kick}} \over 0.5} \right ) \left ( R_{10}/3 \right )
\langle v^2 \rangle_7^{1/2}
$$
when the resultant kick velocity was $100\langle v^2 \rangle_7^{1/2}$km/s.
This gives a modest initial spin period $P_0 \sim 150/v_7$ms.
For a single impulse, the resulting ${\vec \Omega}$ is always orthogonal to 
the space
velocity. Of course for long duration kicks $\tau_K \gg P_0$ the transverse
component of the kick rotationally averages to 0, leading to an aligned spin.

%\begin{figure}[h!] 
%\plottwo{/home/rwr/text/PWN/Tori/ApJ/f2a.eps}{/home/rwr/text/PWN/Tori/ApJ/f2b.eps}
%\plottwo{/home/rwr/text/PWN/Tori/ApJ/f2c.eps}{/home/rwr/text/PWN/Tori/ApJ/f2d.eps}
%\plottwo{/home/rwr/text/PWN/Tori/ApJ/f2e.eps}{/home/rwr/text/PWN/Tori/ApJ/f2f.eps}
%\caption{Figure 1. -- CXO PWNe and best fits torus models (Ng \& Romani 2004). The data
%quality trend from  obvious tori (top row: Crab, Vela) through clear tori 
%(PSR J2229+6114, PSR J1930+1852) to possible tori 
%(bottom row: PSR B1706-44, PSR J0538-2817).}
%%\plotfiddle{file,vsize,rot,hsf,vsf,htrans,vtrans}
%\end{figure}

	More recent treatments of core coupling through collapse (eg.
Heger et al. 2004) do not support the idea of very slow initial spin,
instead suggesting $P_0 \approx 3-10$ms. Such a pre-existing
spin will make rotational averaging of the transverse kick component even
more effective, increasing the tendency to an aligned proper motion.
Lai, Cordes \& Chernoff (2001) have discussed several
physical mechanisms for producing a kick at core collapse. The most important
are the Harrison-Tademaru electromagnetic kick (requiring $P_0$ of a few 
ms), hydrodynamically-driven anisotropy induced kicks 
($\tau_K \sim \tau_{dyn}|_{R\sim 100 {\rm km}} \sim 100$ms)
and magnetic field-induced neutrino anisotropy kicks
($\tau_K \sim \tau_\nu \sim 3$s). They discussed rotational averaging of these
kicks, concluding that the spin-kick orientation could be a significant
constraint on these models. 

	For pulsars born in close 
binaries with aligned angular momenta, the Blaauw mechanism
guarantees a component of the proper motion perpendicular to the (pre-SN)
spin axis. Similarly the binary-like structure of a maximally rotating
core with a strong $m=1$ perturbation can, when the lower mass proto-NS
disrupts, induce a kick to the main core, as
recently discussed by Colpi \& Wasserman (2002). This can be thought of
as an `intra-core Blaauw mechanism' and similarly gives rise to a 
kick component orthogonal to the initial spin. So there are viable models
for both aligned and orthogonal momenta.

	The spectacular CXO images of the Crab and Vela PWNe show clear 
symmetry axes. It was promptly noted that the proper motion vectors
(from 
HST for the Crab and the LBA for Vela; Dodson et al 2003) were roughly aligned. A
more careful assessment (Ng \& Romani 2004) however shows a statistically
significant misalignment; the chance probability of getting two such
2-D projected alignments is $\sim 3$\%. Thus, the alignment can provide
a significant probe of core collapse physics, but more and better measurements
are clearly needed.

\section{CXO measurements of PWNe symmetry axes}

	In Ng \& Romani (2004) we developed a fitting method that can
extract PWN orientations from sparse, Poisson statistics dominated CXO images
by modeling relativistic central tori and jets in pulsar wind nebulae. This
fitting is most sensibly applied to young $\tau_c=10^4\tau_4$y, high
field $B_s=10^{12}B_{12}$G pulsars in 
the high pressure interiors of supernova remnants. There
we will see the axial symmetry of the PWN when the wind termination shock
(torus) scale
$$
r_{ws} \approx  ({\dot E}/4\pi P_{ext})^{1/2} =  
0.17{\rm pc} (B_{12}\tau_4P_{-9})^{-1}
$$
is smaller than the bow shock standoff distance
$$
r_{bs} \approx ({\dot E}/4\pi c \rho_{ext} v^2)^{1/2} =
0.42{\rm pc} (B_{12}\tau_4 v_7)^{-1},
$$
i.e. the pulsar motion must be subsonic. van der Swaluw, et al (2003)
have emphasized that the PWN structure can also be affected when it is 
`crushed' by the SNR reverse shock. Interestingly, since our study of the PWNe 
requires that they be spatial resolved, one selects by angular scale
$\theta_{ws} = r_{ws}/d \propto {\dot E}^{1/2} / d$, 
which is the same scaling
expected for the non-thermal X- and $\gamma$-ray flux.
So these objects are also interesting for study of their magnetospheric
emission and, not coincidentally, study of the PWNe can constrain the
pulsar viewing geometry and aid in the understanding of their high energy 
pulsations.

	The fitting of Ng \& Romani (2004) gives values for the size,
orientation and post-shock flow speed $\beta$ for equatorial tori (sometimes
double) and/or polar jets. Crucially, we also provide simulations that give
statistical errors on the fit parameters. Of course, when the image counts
are large, statistical errors are an underestimate of the true uncertainties
and when the image is particularly sparse, the uniqueness of the model is
questionable. Happily, the unmodeled extra structures seem to have little
effect on the determination of the overall symmetry axis (the key parameter
for the present discussion) and reasonable {\it CXO} exposures can provide 
enough counts to give well exposed images of many sources. For example
in the PSR B1706$-$44 PWN image in figure 1 (lower left) the torus+jet
structure is perhaps less than convincing. Our new {\it CXO} image (Figure 2)
however shows that this interpretation is in good shape -- extended
jets and the bright arc of the near side of the torus are now well seen.
\begin{figure}[t!] 
\plotfiddle{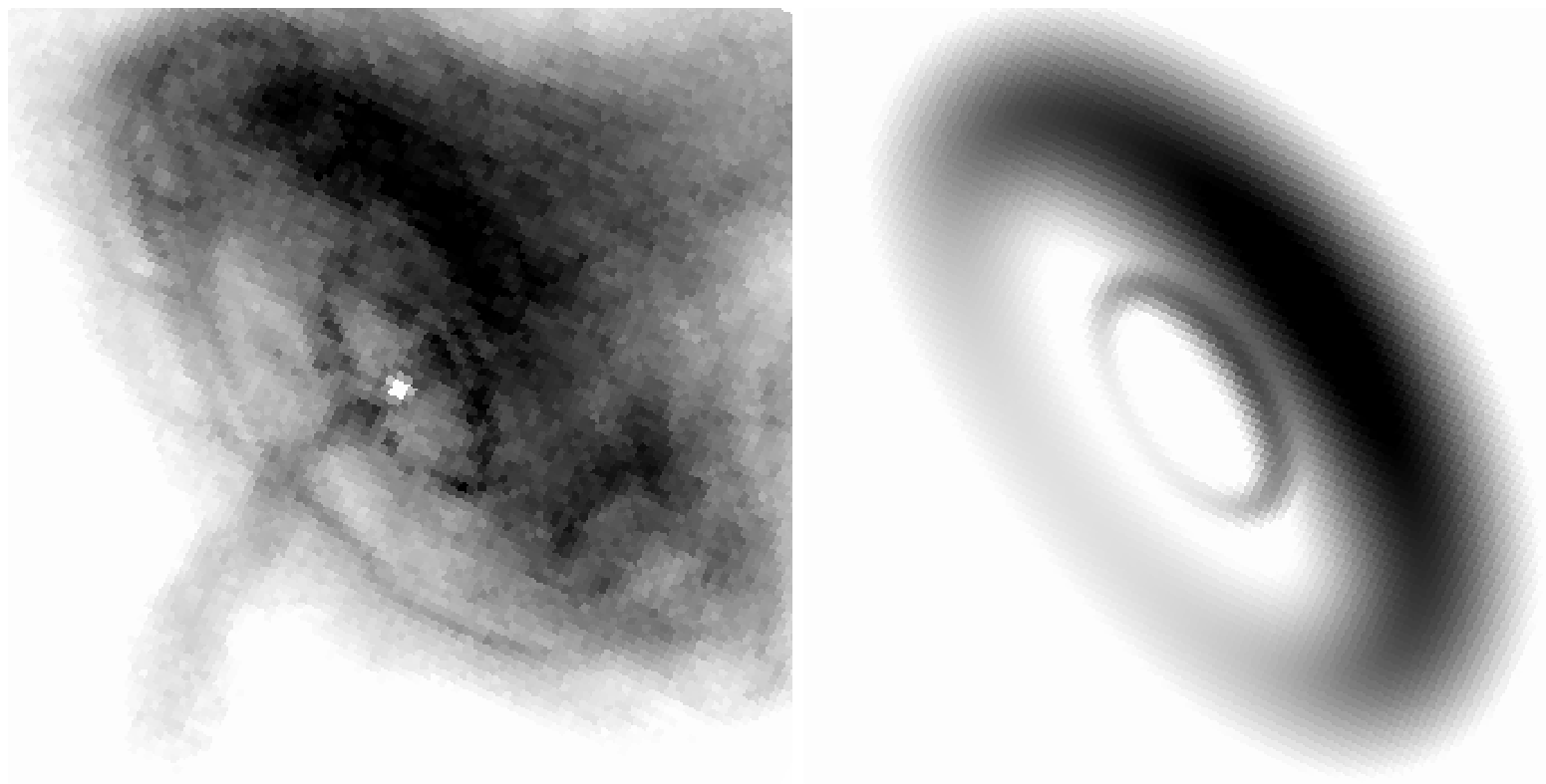}
{3.5truecm}{0}{37}{37}{-183}{-48}
\plotfiddle{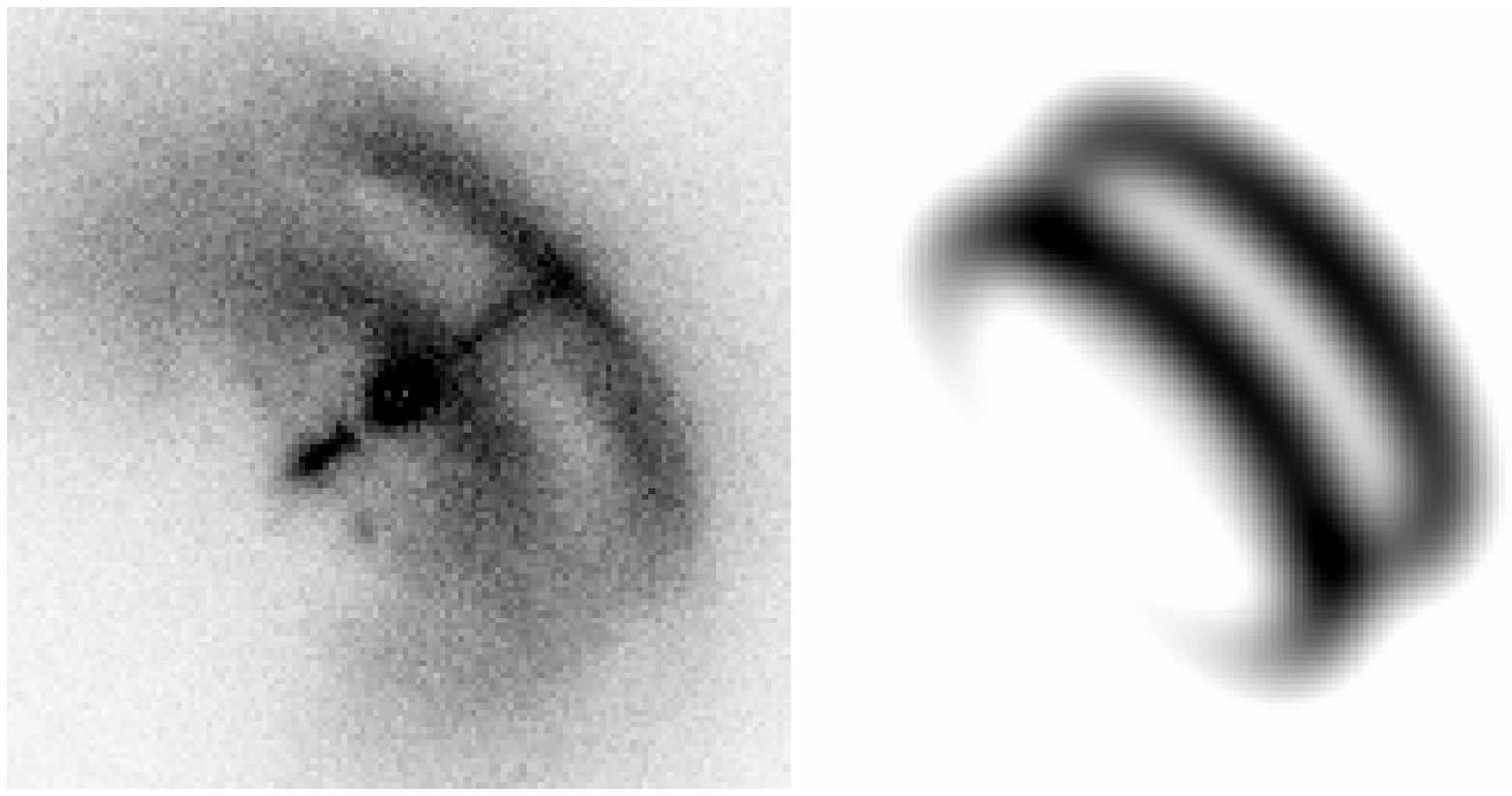}
{0truecm}{0}{33}{33}{-10}{-26}
\plotfiddle{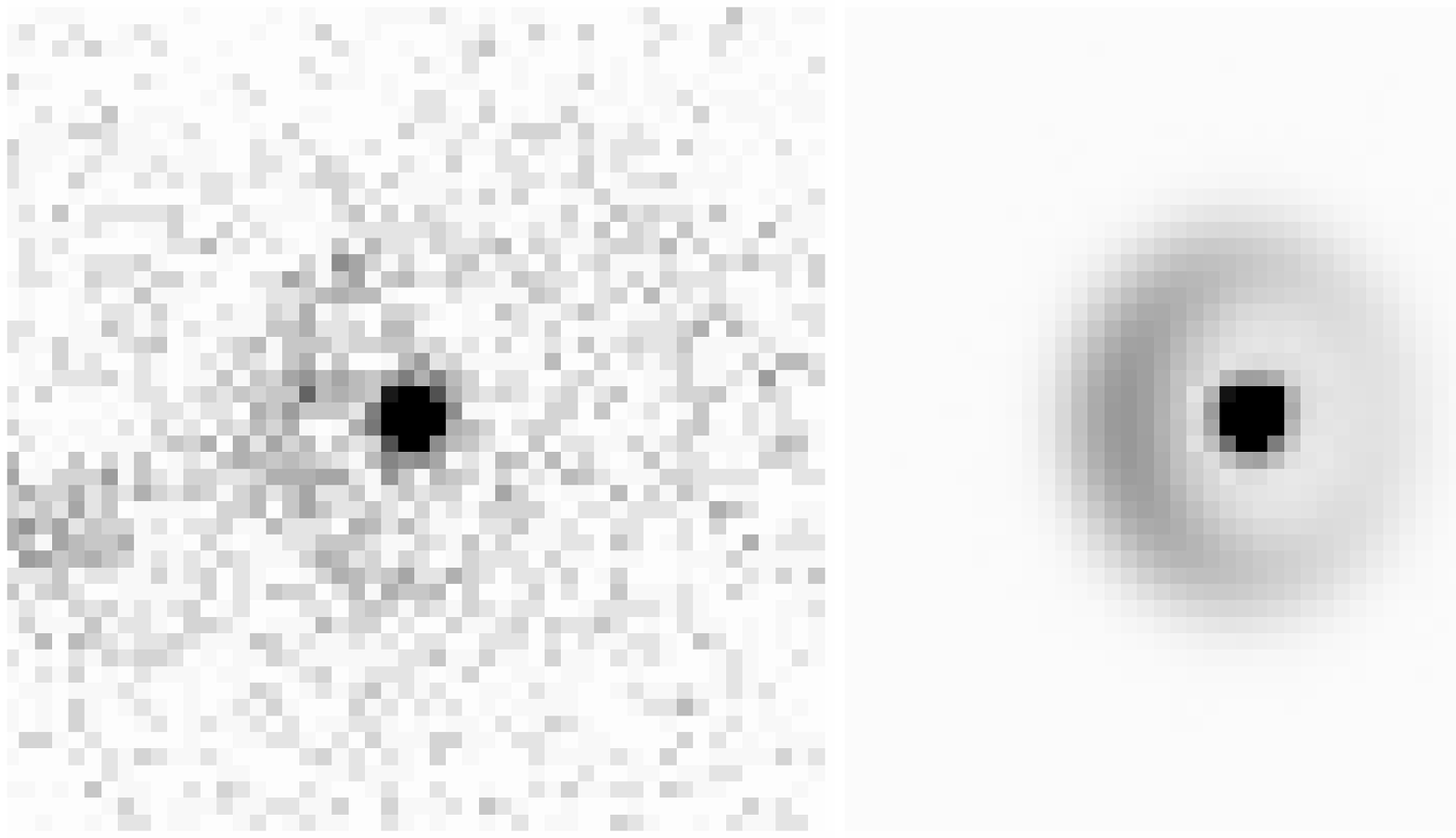}
{0truecm}{0}{33}{33}{-190}{-87}
\plotfiddle{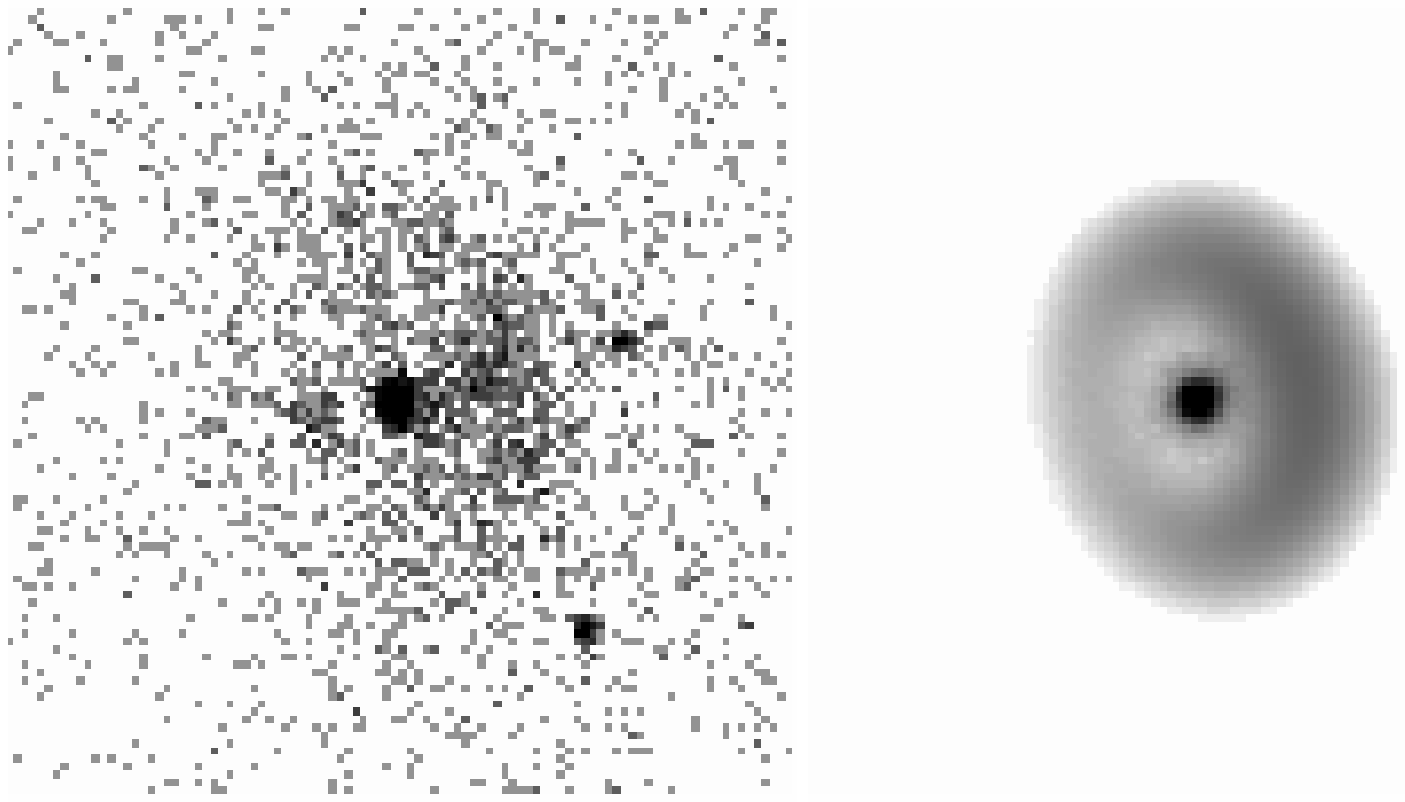}
{0truecm}{0}{37}{37}{+6}{-60}
\plotfiddle{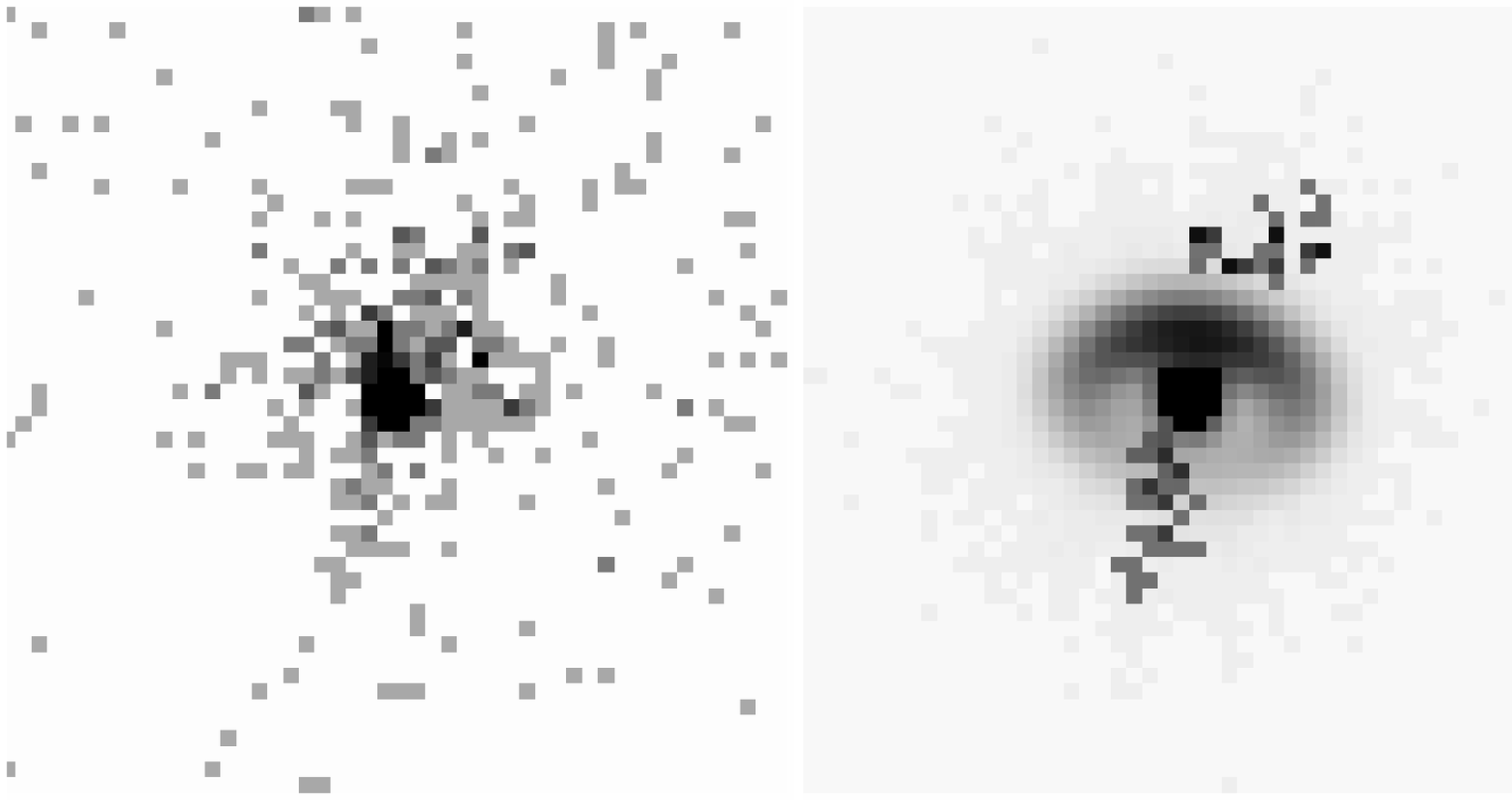}
{0truecm}{0}{33}{33}{-190}{-121}
\plotfiddle{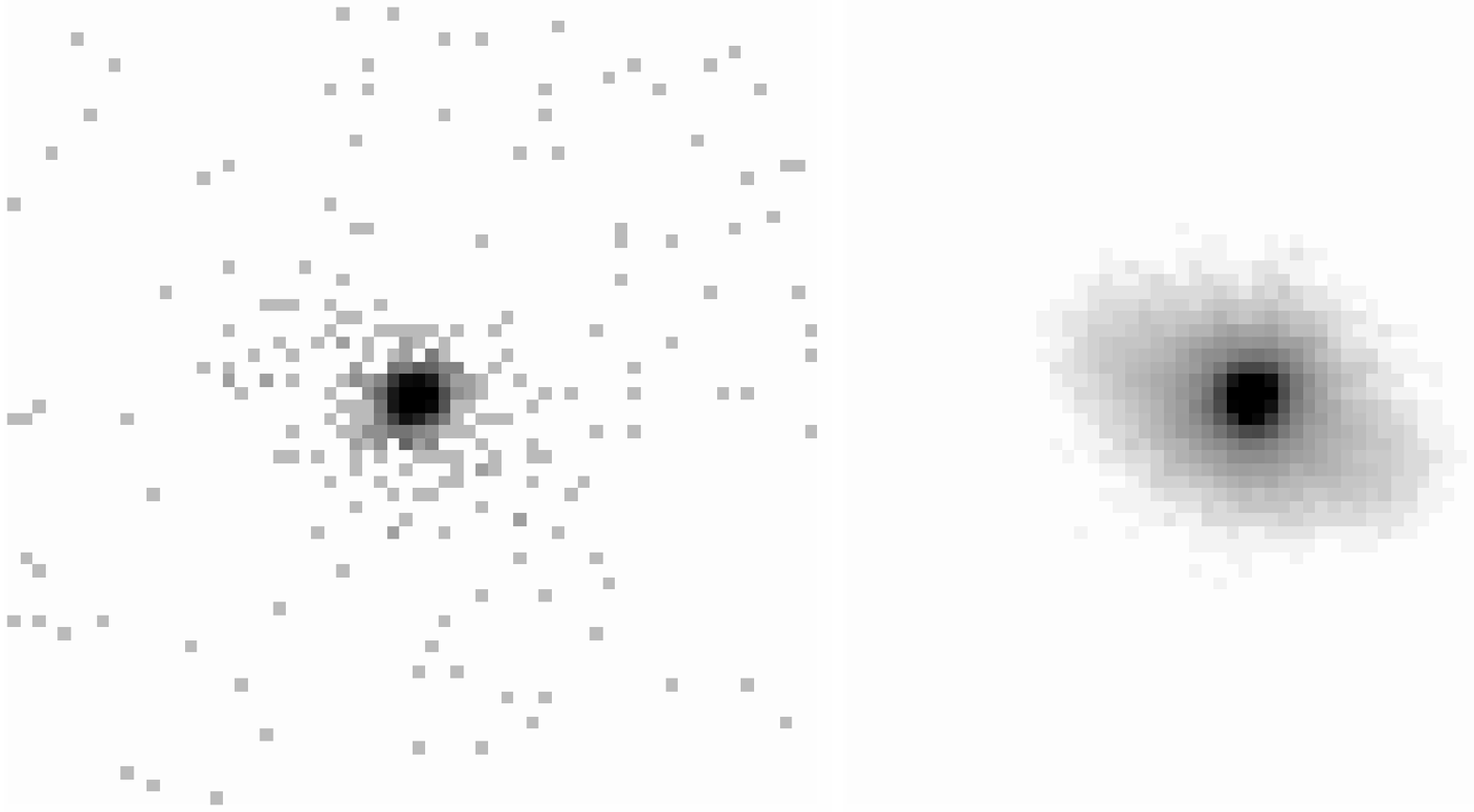}
{0truecm}{0}{33}{33}{-10}{-101}
\caption{CXO PWNe and best-fit torus models (Ng \& Romani 2004). The data
quality trend from  obvious tori (top row: Crab, Vela) through clear tori 
(PSR J2229+6114, PSR J1930+1852) to possible tori 
(bottom row: PSR B1706$-$44, PSR J0538+2817).}
\end{figure}

\section{Spin - Kick Correlation}

	We now wish to compare the PWN-measured spin axes with the kick vectors.
Establishing ${\vec v}$ is difficult and we must rely on a variety of methods.
Optical or radio interferometric proper motions are of course best,
and these are becoming available for several young objects. When we lack
direct proper motions, we can often make estimates from the offset from
the birthsite, since as discussed above toroidal PWNe will almost invariably 
be inside their parent SNR. This method is limited by the accuracy with which
the explosion center can be measured, and we will always prefer direct 
measurements. For example, the offset of PSR J0538+2817 from the center of S147
was used by Romani \& Ng (2003) to estimate the proper motion direction. This was
supported by a timing proper motion (Kramer et al 2003), which solidified
the association with the SNR and gave a more precise pulsar age. However,
a precise position angle is still required and with Walter Brisken (NRAO)
we have a program underway to measure this.

\begin{figure}[b!]
\plotfiddle{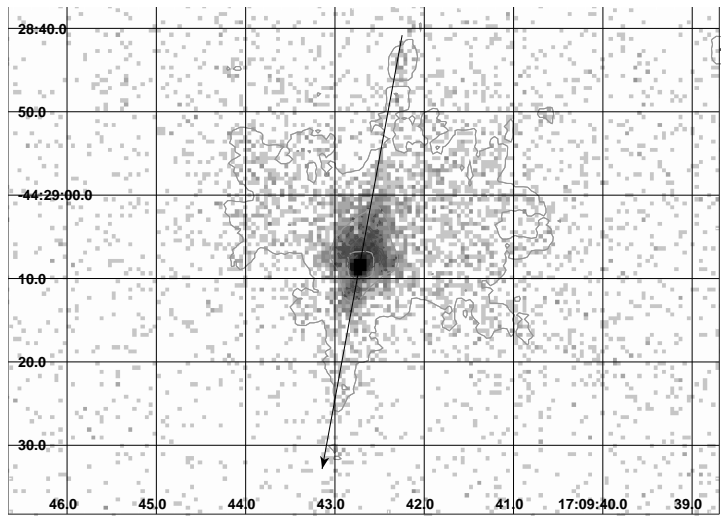}
{5.9truecm}{0}{50}{50}{-150}{-105}
\caption{A raw 0.5-7keV image of the PWN of PSR B1706$-$44 from
a follow-on 100ks {\it CXO} pointing. Such moderately deep exposures
can make the PWN PA fitting quite robust.}
\end{figure}

	In Ng \& Romani (2004) we described six proper motion-spin axis
comparisons. The Crab and Vela pulsars are well known (although the 
substantial errors are not always appreciated). For PSR B1951+32 we compared
the interferometric proper motion (Migliazzo et al 2002) with the spin axis
PA fit from the optical jets. For PSRs B1706$-$44 and J0538+2817 we compare
(for now) with offsets from the SNR centers. These last measurements 
have at present limited accuracy, but we are working toward improved {\it CXO}
imaging and precision proper motions for these objects. For PSR B0656+14
we show that the spin axis poiinted nearly at Earth is consistent with the small
interferometric proper motion (Brisken et al 2003), but unfortunately, the
PWN appears to have a surface brightness too low for an accurate independent
spin axis position angle. We mention one additional alignment here -- 
PSR J1124$-$5916 in 
G292.0+1.8 shows a clear offset from its nearly circular SNR center. We have
measured the elongation of the central PWN and compared this with the direction
to the explosion center as determined from the radio image of the forward shock 
(Gaensler \& Wallace 2004). The resulting angle 
$\theta_{\Omega-v} \approx 22\pm 7^\circ$ shows substantial mis-alignment
if the PWN major axis is identified with the polar jets.  Unfortunately better 
imaging is needed here to make the jet interpretation secure and
so at present $\theta_{\Omega-v}$ has a $\pi/2$ ambiguity for this source.

	Now we can compare these angles with other pulsar parameters
to constrain the kick physics. With detailed modeling it is interesting
to compare with the amplitude of the proper motion (Ng \& Romani
in preparation), but for now we describe only the simplest comparison:
that with the initial spin $P_0$. Estimating $P_0$ is itself non-trivial
and generally requires a kinematic age and some constraint on the effective
braking index. Several useful estimates are in Migliazzo et al (2002); others
can be made. For Vela, we can for example use the measured $n=1.4\pm 0.2$
and the kinematic age (dominated by the explosion center uncertainty) to
get $P_0=13\pm13$ms, which is of some use. However, pulsars 
with large $P_0$ are of the greatest interest. For PSR J1124-5916
there remains some uncertainty in the distance. Combined with the braking
index uncertainty, we derive $P_0 = 78\pm 24$ms. PSR J0538+2817 is, on
this score, truly outstanding as, with a kinematic age $\ll \tau_c$, it
must have $P_0$ very close to its present 143ms period.

\begin{figure}
\plotfiddle{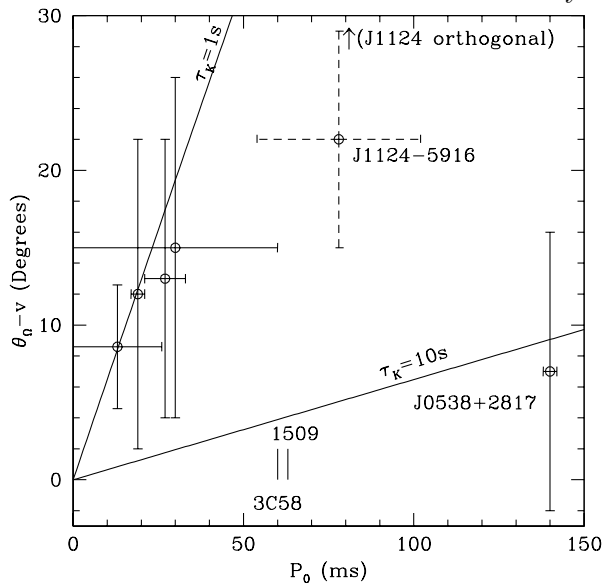}
{6.4truecm}{0}{40}{40}{-120}{-70}
\caption{Spin-Kick angles {\it vs.} estimated initial spin periods.
The lower limit to the residual alignment after rotational averaging 
is shown for two characteristic kick durations. The value for PSR J1124$-$5916
is shown dashed, since the orthogonal solution still remains viable. The
$P_0$ of two other pulsars with slow initial spins, but no proper motion
estimates presently available, are also shown.}
\end{figure}

	The first thing that we infer from these data is that there is a
true causal correlation between the spin and kick position angles. Even
ignoring PSR J1124, we find a chance probability of $4 \times 10^{-4}$ 
that the projections of the kick and spin angles are aligned (2-D) 
within the 1$\sigma$ upper limits 
(the probability of getting a set of angles as small
as the best fits is $2 \times 10^{-5}$). However, we also infer that the
kicks are significantly misaligned -- the mean offset is $10^\circ$,
a $4\sigma$ difference from 0. Finally we see that there is a general trend
toward poorer alignment at large $P_0$. This trend is consistent with the
residual misalignment from a few second kick. This is 
the timescale for momentum imparted by anisotropic $\nu$ emission
during the quasi-static core cooling phase, which seems fairly reasonable.
The exception is PSR J0538+2817, which requires $\tau_K$ of 10s or
more, rather difficult to reconcile with neutrino cooling times.

	This trend to alignment must be contrasted with the model of 
PSR B1913+16 by Wex, Kalogera \& Kramer (2000), which shows that for initially
aligned spins, the second pulsar was kicked at $\theta_{\Omega-v}=80\pm5^\circ$.
Should we infer that natal kicks are aligned for single stars, orthogonal
for binaries? Not necessarily. Clearly the precession of the PSR/B star
binaries (e.g. Kaspi, these proceedings) implies a kick component out of
the original orbital plane. The orbit and scintillation velocity data
for PSR J0737$-$3039 (see Willems \& Kalogera 2004; Ransom, these proceedings)
also indicate a large out-of-plane kick. Detection and calibration of the 
geodetic precession cone angle should give a uniquely precise measurement of this
pulsar's
kick direction. When we recall that binary survival puts a strong selection
bias toward (retrograde) kicks in the orbit plane, it seems likely that
a trend toward kick alignment can be present in binaries, as well.
\medskip

	This is work in progress, but at the moment a substantial, but
incomplete kick alignment seems present in most young pulsars. The degree
of alignment suggests kicks lasting a few seconds, so
a neutrino mediated kick seems tenable. As discussed by
Lai {\it et al.} (2001), the most plausible mechanisms for producing a 
long-lived anisotropy invoke large $\ge 10^{15}$G organized fields in the 
proto-NS interior. However, a few of the larger $P_0$ pulsars are a challenge
for this picture. If good alignment persists for these, some sort of
post-collapse momentum kick, such as the super-Eddington accretion/asymmetric
jet picture suggested in Romani \& Ng (2003) may be required.
Even if $\tau \sim 3$s neutrino kicks dominate, rotational averaging has 
serious implications for the survival of pulsar binaries, since in-plane
kick components should be greatly reduced. Thus further observation and modeling
to constrain the {\it vector} properties of neutron star kicks seems essential
to understand both kick physics and the pulsar population.
\medskip

Supported in part by NASA grants SAO GO2-3085X and NAGS-13344.

\end{document}